\documentclass[aps,prx,twocolumn,notitlepage,showkeys,showpacs,superscriptaddress,longbibliography,footinbib, preprintnumbers]{revtex4-2}

\usepackage{epsf}
\usepackage{here,times}
\usepackage[colorlinks=true,citecolor=blue,linkcolor=blue, urlcolor=blue]{hyperref} 
\usepackage{slashed}
\usepackage{amsthm}
\usepackage{thmtools}
\usepackage{blindtext}
\usepackage{physics, bm}
\usepackage{mathrsfs}
\usepackage{amsfonts, mathtools, amsmath,amssymb,dsfont}
\usepackage{esvect}
\usepackage{enumerate,dcolumn,multirow,bbold, enumitem}
\usepackage{longtable}
\usepackage{array}
\usepackage{tikz-cd}
\usepackage{bbding}
\usepackage[normalem]{ulem}
\usepackage{mathrsfs}
\usepackage{calligra}
\usepackage{multirow}
\usepackage{bbm}
\usepackage{mathbbol}

\usepackage{algorithm, algorithmic}

\declaretheoremstyle[
shaded={bgcolor=\color{rgb}{0.9,0.9,0.9}}  % comment this line in/out
]{theorem}

\declaretheoremstyle[
shaded={bgcolor=\color{rgb}{0.9,0.9,0.9}}% comment this line in/out
]{question}

\declaretheoremstyle[
shaded={bgcolor=\color{rgb}{0.9,0.9,0.9}}  % comment this line in/out
]{remark}

\declaretheoremstyle[
shaded={bgcolor=\color{rgb}{0.9,0.9,0.9}}  % comment this line in/out
]{proposition}

\declaretheoremstyle[
shaded={bgcolor=\color{rgb}{0.9,0.9,0.9}}  % comment this line in/out
]{definition}

\declaretheoremstyle[
shaded={bgcolor=\color{rgb}{0.9,0.9,0.9}}  % comment this line in/out
]{assumption}

\declaretheoremstyle[
shaded={bgcolor=\color{rgb}{0.9,0.9,0.9}}  % comment this line in/out
]{conjecture}

\declaretheoremstyle[
shaded={bgcolor=\color{rgb}{0.9,0.9,0.9}}  % comment this line in/out
]{corrorary}

\declaretheoremstyle[
shaded={bgcolor=\color{rgb}{0.9,0.9,0.9}}  % comment this line in/out
]{axiom}

\declaretheoremstyle[
shaded={bgcolor=\color{rgb}{0.9,0.9,0.9}}  % comment this line in/out
]{lemma}

\declaretheoremstyle[
shaded={bgcolor=\color{rgb}{0.9,0.9,0.9}}  % comment this line in/out
]{problem}

\usepackage{epsf}
\usepackage{here,times}
\usepackage{slashed}
\usepackage{amsthm}
\usepackage{thmtools}
\usepackage{blindtext}
\usepackage{physics, bm}
\usepackage{mathrsfs}
\usepackage{amsfonts, mathtools, amsmath,amssymb,dsfont}
\usepackage{esvect}
\usepackage{enumerate,dcolumn,multirow,bbold, enumitem}
\usepackage{longtable}
\usepackage{array}
\usepackage{tikz-cd}
\usepackage{bbding}
\usepackage[normalem]{ulem}
\usepackage{mathrsfs}
\usepackage{calligra}

\newcommand{\RNum}[1]{\uppercase\expandafter{\romannumeral #1\relax}}

\newcommand{\Rom}[1]{ \uppercase\expandafter{\romannumeral#1}}

\newcommand{\kk}{{\bf{{k}}}}

\newcommand{\HBdG}{\mathcal{H}_{\text{BdG}}}
\newcommand{\Deltasc}{\Delta_{\text{sc}}}

\definecolor{ZZYcolor}{rgb}{0.1,0.5,0.4}

\begin{document}
\title{Double Majorana Vortex Flat Bands in the Topological Dirac Superconductor}
\author{Zhongyi Zhang}
\affiliation{Department of Physics, Hong Kong University of Science and Technology, Clear Water Bay, Hong Kong, China}

\author{Zixi Fang}
\affiliation{Beijing National Laboratory for Condensed Matter Physics, and Institute of Physics,
Chinese Academy of Sciences, Beijing 100190, China}
\affiliation{School of Physical Sciences, University of Chinese Academy of Sciences, Beijing 100049, China}

\author{Shengshan Qin}
\affiliation{School of Physics, Beijing Institute of Technology, 100081 Beijing, China}

\author{Peng Zhang}
\affiliation{School of Physics and National Laboratory of Solid State Microstructures, Nanjing University, Nanjing, China}

\author{Hoi Chun Po}
\thanks{Corresponding authors:~\href{mailto:hcpo@ust.hk}{hcpo@ust.hk}}
\affiliation{Department of Physics, Hong Kong University of Science and Technology, Clear Water Bay, Hong Kong, China}

\author{Xianxin Wu}
\thanks{Corresponding authors:~\href{mailto:xxwu@itp.ac.cn}{xxwu@itp.ac.cn}}
\affiliation{CAS Key Laboratory of Theoretical Physics, Institute of Theoretical Physics, Chinese Academy of Sciences, Beijing, China}

\begin{abstract}

Vortex lines, known as topological defects, are cable of trapping Majorana modes in superconducting topological materials. Previous studies have primarily focused on topological bands with conventional $s$-wave pairing. However, topological Dirac semimetals exhibiting a unique orbital texture can favor unconventional pairing when electronic correlations are significant. The topology of vortices in these systems has yet to be explored. In this work, we investigate the vortex bound states in superconducting Dirac semimetals, with a particular focus on the orbital-singlet unconventional pairing, which generates higher-order Majorana hinge modes. Remarkably, we identify robust double Majorana vortex flat bands at zero energy. In type-I Dirac semimetals, these Majorana flat bands are located between the projections of two superconduting Dirac points. In contrast, in type-II Dirac semimetals, they extend across the entire 1D Brillouin zone. These double flat bands arise from a nontrivial $\mathbbm{Z}_2$ topology defined by an effective particle-hole symmetry and are protected by the four-fold rotational symmetry. Additionally, we observe that moving the vortex line close to a hinge can trivialize the higher-order Majorana arc on the hinge, leaving a single Majorana mode at the vortex core due to the hybridization of Majorana modes. Finally, we discuss the potential experimental implications for correlated Dirac semimetals, such as electron-doped iron-based superconductors.

\end{abstract}
\maketitle

\section{Introduction}

When nontrivial band topology meets superconductivity, it paves the way for the emergence of fascinating topological superconductivity, even without the necessity for exotic pairing states~\cite{RevModPhys.83.1057,PhysRevLett.100.096407,alicea2012new,PhysRevLett.105.077001,beenakker2013search,PhysRevLett.104.040502,zhang2024fermi}. This intersection can give rise to Majorana zero modes (MZMs) at surfaces, edges, or within vortices~\cite{sato2017topological,PhysRevB.81.125318,zhang2024topological,wong2023higher,PhysRevLett.120.067003,PhysRevResearch.3.013066,PhysRevB.106.L121108}. The non-Abelian braiding properties of MZMs endow them with inherent stability against local perturbations, highlighting their significant potential for applications in topological quantum computation~\cite{kitaev2001unpaired,RevModPhys.80.1083,sarma2015majorana,RevModPhys.87.137}. The realm of nontrivial band topology includes phases such as topological insulators~\cite{sato2011unexpected,xu2011topological,PhysRevB.76.205304}, Dirac semimetals~\cite{yang2014classification,PhysRevB.88.125427,PhysRevB.85.195320}, and Weyl semimetals~\cite{yan2017topological,PhysRevX.5.031013,soluyanov2015type}. To integrate the two essential ingredients together for realizing MZMs, tremendous experimental efforts are devoted to developing heterostructures consisting of topological insulators and conventional superconductors, where topological surface states acquire a superconducting gap by the proximity effect~\cite{PhysRevLett.107.217001,wang2018evidence,PhysRevLett.114.017001,kong2019half,PhysRevLett.116.257003,wang2012coexistence,liu2024signatures}. However, challenges such as complex interfaces and low working temperatures often hinder the detection and manipulation of MZMs in such systems. This has led to the exploration of integrating topological properties and superconductivity within a single material system. High-T$_c$ iron-based superconductors (IBS) have emerged as such an ideal platform, as theoretical~\cite{hao2014topological,PhysRevB.93.115129,PhysRevB.92.115119,PhysRevLett.117.047001,hao2019topological,wu2022pursuit} and experimental~\cite{zhang2018observation,zhang2019multiple} studies have identified their intrinsic topological insulating characteristics. Consequently, zero-bias peaks observed in vortex cores across various IBS provide compelling evidence for vortex MZMs~\cite{nadj2014observation,vaitiekenas2020flux,machida2019zero,PhysRevX.8.041056}.

In contrast to topological insulators, topological Dirac semimetals are characterized by bulk Dirac points protected by time-reversal and crystalline symmetries~\cite{yang2014classification,PhysRevB.88.125427,PhysRevB.85.195320}. These semimetals exhibit side Fermi surface arcs, which, unlike those in Weyl semimetals~\cite{jia2016weyl,hasan2017discovery,xu2015discovery,huang2015weyl}, are not topologically protected~\cite{kargarian2016surface,le2018dirac}. Instead, they display stable high-order Fermi arcs localized on hinges~\cite{wieder2020strong}. The metallic bulk states of Dirac semimetals can support superconductivity, and indeed, several Dirac materials, such as Cd$_3$As$_2$~\cite{wang2016observation,aggarwal2016unconventional,he2016pressure} and Au$_2$Pb~\cite{xing2016superconductivity}, have been found to exhibit intrinsic superconductivity. Moreover, IBS are known to host topological bulk Dirac cones, which can be achieved through electron doping~\cite{zhang2019multiple}. The interplay between superconductivity and the topological semimetal phase can generate intriguing topological states~\cite{PhysRevLett.115.187001}. With $s$-wave pairing, a topological nodal vortex can be realized, which can be gapped to support MZMs through crystalline symmetry breaking~\cite{PhysRevLett.123.027003,PhysRevLett.122.207001,PhysRevLett.129.277001}. Furthermore, when electronic correlation effects, particularly relevant for iron-based superconductors, are considered, Dirac semimetals can favor unconventional pairing due to the unique orbital texture on the 3D Fermi pockets~\cite{PhysRevLett.115.187001,PhysRevB.94.014510}. This pairing state can result in multi-fold surface Majorana modes and even high-order Majorana hinge flat bands~\cite{PhysRevB.106.214510,xie2024hinge}. It is known that the energy spectra of vortices is intimately related to the intrinsic topology of systems. While prior studies have primarily focused on conventional $s$-wave pairing~\cite{PhysRevLett.129.277001,qin2019topological,PhysRevLett.124.257001,zhang2023gapless}, there is a scarcity of systematic investigations into the vortex bound states of unconventional pairing states. In correlated Dirac semimetals, such as electron-doped IBS, the topology of vortices with unconventional pairing and their interplay with higher-order hinge states remain to be comprehensively understood.

Motivated by this, we study the vortex bound states in superconducting  Dirac semimetals with the time-reversal symmetry, placing particular emphasis on orbital-singlet unconventional pairing that emerge due to the orbit texture associated with Dirac points.
Remarkably, we identify robust double Majorana vortex flat band at zero energy. In type-I Dirac semimetals, these Majorana flat bands are situated in the region between the projections of two Bogoliubov-de Gennes Dirac points. In contrast, in type-II Dirac semimetals, they extend across the entire 1D Brillouin zone. 
To understand their origin, we develop the bulk-vortex correspondence for such double-degenerate vortex flat bands and confirmed their topological $\mathbbm{Z}_2$ nature defined by an effective particle-hole symmetry. These Majorana flat bands carry different angular momenta and thus are protected by the four-fold rotational symmetry. Due to the Dirac semimetal with orbital-singlet pairing in the spin subspace exhibiting characteristics similar to Rashba electron gas with conventional $s$-wave pairing, a double degenerate Majorana zero modes is realized.
Furthermore, we explore the interplay between the vortex Majorana flat bands and the higher-order Majorana arc localized along the hinge. By moving the vortex line close to one hinge, we observe that the hybridization between Majorana modes can trivialize the higher-order Majorana arc on the hinge and leave a single Majorana mode at the vortex core, which is robust against lattice symmetry breaking.
Finally, we discuss the potential experimental implications for correlated Dirac semimetals, such as doped IBS.

\section{Topological Dirac Superconductor}
\begin{figure}[t]
	\begin{center}
		\includegraphics[width=0.99\columnwidth]{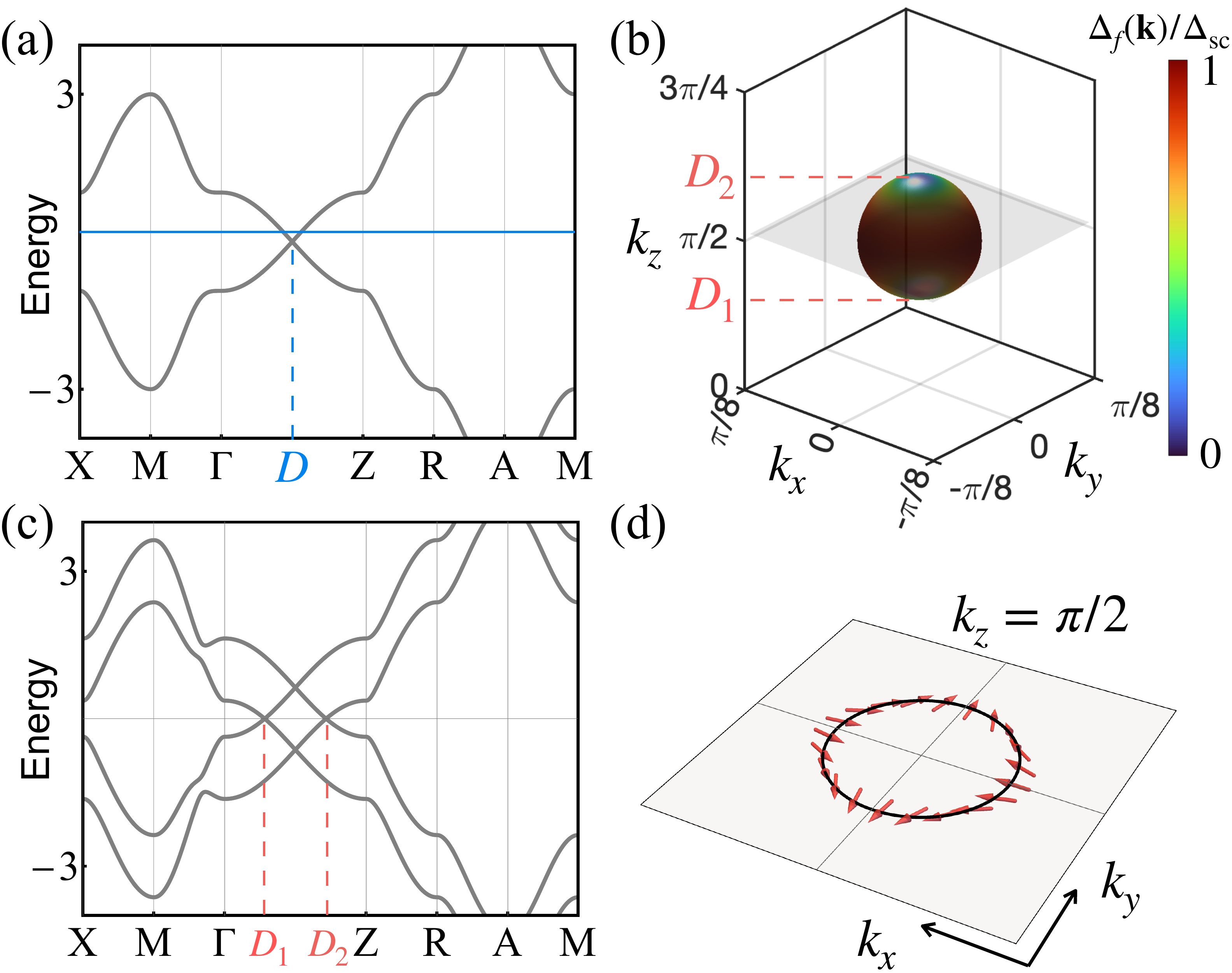}
		\caption{\label{fig_band}
        (a) The band structure of normal states described by Eq.~\eqref{eq_normal}, with
        $\Gamma$, $\mathrm{X}$, $\mathrm{M}$, $\mathrm{Z}$, $\mathrm{R}$, $\mathrm{A}$ representing $(0,0,0)$, $(\pi,0,0)$, $(\pi,\pi,0)$, $(0,0,\pi)$, $(\pi,0,\pi)$, $(\pi,\pi,\pi)$ in the Brillouin zone, respectively.
        The blue solid line represents the chemical potential.
        $D$ represents the momentum at which the Dirac point appears.
        (b) shows the Fermi surface corresponding to the Fermi energy illustrated in (a), with the color representing the magnitude of the superconducting gap projected onto the Fermi surface.
        (c) The band structure of BdG Hamiltonian described by Eq.~\eqref{eq_bdg}.
        $D_{1,2}$ represents the momentum at which the BdG Dirac point appears.
        (d) Illustration of the $\mathbf{d}$-vector at the intersections of the Fermi surface with the $k_z=\pi/2$ plane in the band basis.
        The parameters are set to be $\{m,t,t_z,\eta,\lambda_1,\lambda_2,\mu\}=\{2,1,1,1,-1,2,0.2\}$.
        }
	\end{center}
\end{figure}

We begin with the general effective Hamiltonian for topological Dirac semimetals in tetragonal systems, where Dirac points are protected by lattice and time-reversal symmetries encoded in the magnetic space group $\mathcal{G} = P4/mmm1^\prime$~\cite{PhysRevLett.115.187001}. 
 We choose a minimum set of bases at the $\Gamma$ point characterized by different angular momentum $J_z$ and parity $p$: ${|J_z,p\rangle=|\pm
\frac{1}{2},p \rangle, |\pm\frac{3}{2},\bar{p} \rangle}$, where $p = +,-$ and $\bar{p}$ is the opposite of $p$.
The band inversion between them either around the $\Gamma$ or $\mathrm{Z}$ point leads to nontrivial band topology. In conventional Dirac semimetals, the $J_z=\pm \frac{1}{2}$ basis corresponds to $s$ orbital even-parity states and $J_z=\pm \frac{3}{2}$ basis corresponds to $p_{x,y}$ orbital odd-parity states. In contrast, within IBS the $J_z=\pm \frac{1}{2}$ basis is odd-parity and corresponds to As/Se $p_z$ orbitals coupled with Fe $d_{xy}$ orbital, whereas the $J_z=\pm \frac{3}{2}$ basis is even-parity and corresponds to the Fe $d_{xz/yz}$ orbitals coupled with As/Se $p_{x/y}$ orbital~\cite{hao2019topological,wu2022pursuit}.
Within these bases, the generators of the symmetry group $\mathcal{G}$, including inversion, four-fold rotation and mirror reflection with respect to the $yz$ plane, are represented by,
\begin{align}\label{eq_symmRepI}
 &\mathcal{I}=\sigma_z,\ C_4=is_z\exp[i\frac{\pi}{4}\sigma_zs_z],\ \mathcal{M}_x=is_x.
\end{align}
Here, the Pauli matrices $\sigma_i$ and $s_i$ ($i = 0, 1, 2, 3$) represent the orbital and spin degrees of freedom, respectively.
The time-reversal symmetry is expressed as $\mathcal{T} = i s_y \mathcal{K}$, where $\mathcal{K}$ denotes the complex conjugation operation. The tight-binding Hamiltonian, which preserves the symmetries described above and contains a pair of Dirac points, can be written as
\begin{equation}\label{eq_normal}
\begin{split}
\mathcal{H}_0(\kk)=&\ (m-t\cos k_x-t\cos k_y-t_z\cos k_z)\sigma_z s_0\\
&+\eta (\sin k_x\sigma_x s_z-\sin k_y \sigma_y s_0)\\
 &+\lambda_1\sin k_z(\cos k_y-\cos k_x)\sigma_x s_x\\
 &+\lambda_2\sin k_x\sin k_y \sin k_z\sigma_x s_y.
\end{split}
\end{equation}
The band structure is shown in Fig.~\ref{fig_band}(a), where a pair of Dirac points appears along the $\Gamma$-Z line, located at $(0,0,\pm D)$ with $D=\text{cos}^{-1}\frac{m-2t}{t_z}$. Here, two branches of the Dirac cone have the opposite slope along $k_z$ axis, corresponding type-I Dirac semimetal. In the following sections, we will introduce a tilting term to achieve a type-II Dirac semimetal phase, where two bands of the Dirac cone have slopes of the same sign. The chemical potential $\mu$ is set near the Dirac points, as indicated in Fig.~\ref{fig_band}(a), resulting Fermi surfaces are two 3D pockets around Dirac points, shown in Fig.~\ref{fig_band}(b). In the case, superconductivity can merge and the pairing symmetry depends on the interactions. While an intraorbital attraction favor an intraorbital $s$-wave pairing, an interorbital attraction promotes an orbital-singlet spin-triplet $B_{1u/2u}$ pairing due to the uniqe orbital texture associated with Dirac cones~\cite{PhysRevLett.115.187001,PhysRevB.94.014510}. Given studies on topological properties with $s$-wave pairing, we will focus on this unconventional pairing state and explore the corresponding vortex bound states. 

We consider the orbital-singlet spin-triplet pairing in the $B_{1u}$ channel and the corresponding Bogoliubov-de Gennes (BdG) Hamiltonian is then given by,
\begin{equation}\label{eq_bdg}
    \mathcal{H}_{\text{BdG}}(\kk)=[\mathcal{H}_0(\kk)-\mu]\tau_3+\Delta_{\text{sc}}\sigma_ys_y\tau_1
\end{equation}
with the bases being $\Psi_\kk=(\hat{c}_\kk,-i\sigma_0s_y\hat{c}_{-\kk}^\dagger)$. Here the $\tau_i$ represent the Pauli matrices in Nambu space and $\Delta_{\text{sc}}$ denotes the superconducting order parameter. In the BdG bases, the representation matrices of the three generators of $\mathcal{G}$ are given by,
\begin{align}\label{eq_bdgsymmRepI}
 &\tilde{\mathcal{I}}=\tau_z\sigma_z,\ \tilde{C}_{4z}=i\tau_zs_z\exp[i\frac{\pi}{4}\sigma_zs_z],\ \tilde{\mathcal{M}}_x=i\tau_zs_x.
\end{align}
The time-reversal and particle-hole symmetries are represented as $\tilde{\mathcal{T}} = is_y \mathcal{K}$ and $\tilde{\mathcal{P}} = \tau_y s_y \mathcal{K}$, respectively. The BdG band structure is shown in Fig.~\ref{fig_band}(c), where two pairs of Dirac points along the $\Gamma$-Z line are located in $(0,0,\pm D_1)$ and $(0,0,\pm D_2)$. The superconducting gap magnitude projected onto the Fermi surface is shown in Fig.~\ref{fig_band}(b), where the BdG nodal points are clearly visible at the poles of the Fermi surface. The node of pairing on $k_z$ axis is attributed to orbit-momentum locking, where the inter-orbital pairing vanishes on this axis~\cite{PhysRevB.94.014510}. The existence of the BdG Dirac points is guaranteed by the nonzero winding number $w(k_z)$ of the in-plane normalized $\hat{\mathbf{d}}(k;k_z)$-vector 
\begin{equation}
    w(k_z)=\frac{1}{2\pi}\int_{\text{FS with fixed}\ k_z} \hat{\mathbf{d}}(k;k_z)\times\frac{\dd \hat{\mathbf{d}}(k;k_z)}{\dd k}\ \dd k,
\end{equation}
at a fixed $k_z$ plane~\cite{PhysRevLett.115.187001}, as shown in Fig.~\ref{fig_band}(d).
Here, $\hat{\mathbf{d}}(k;k_z)$ is the $\mathbf{d}$-vector of the superconducting order parameter projected onto Fermi surface, of which expression is
\begin{equation}
    \widetilde{\Delta}_f(\kk)=(\Phi_\kk^f)^\dagger\Deltasc \sigma_y s_y (\Phi_{-\kk}^f)^*=\Delta_f \hat{\mathbf{d}}(\kk)\cdot \mathbf{s}
\end{equation}
where $\Phi_\kk^f$ is the eigenvectors of the normal state Hamiltonian Eq.~\eqref{eq_normal} with eigenvalue $\mu$.
Note that the two BdG Dirac points located at $\pm D_1$ and $\pm D_2$ cannot annihilate with each other, as they carry the same winding number of $\hat{\mathbf{d}}$.

\section{Vortex Majorana Flat Band}
\subsection{Type-I Dirac semiemtal}
\begin{figure}[t]
	\begin{center}
		\includegraphics[width=0.99\columnwidth]{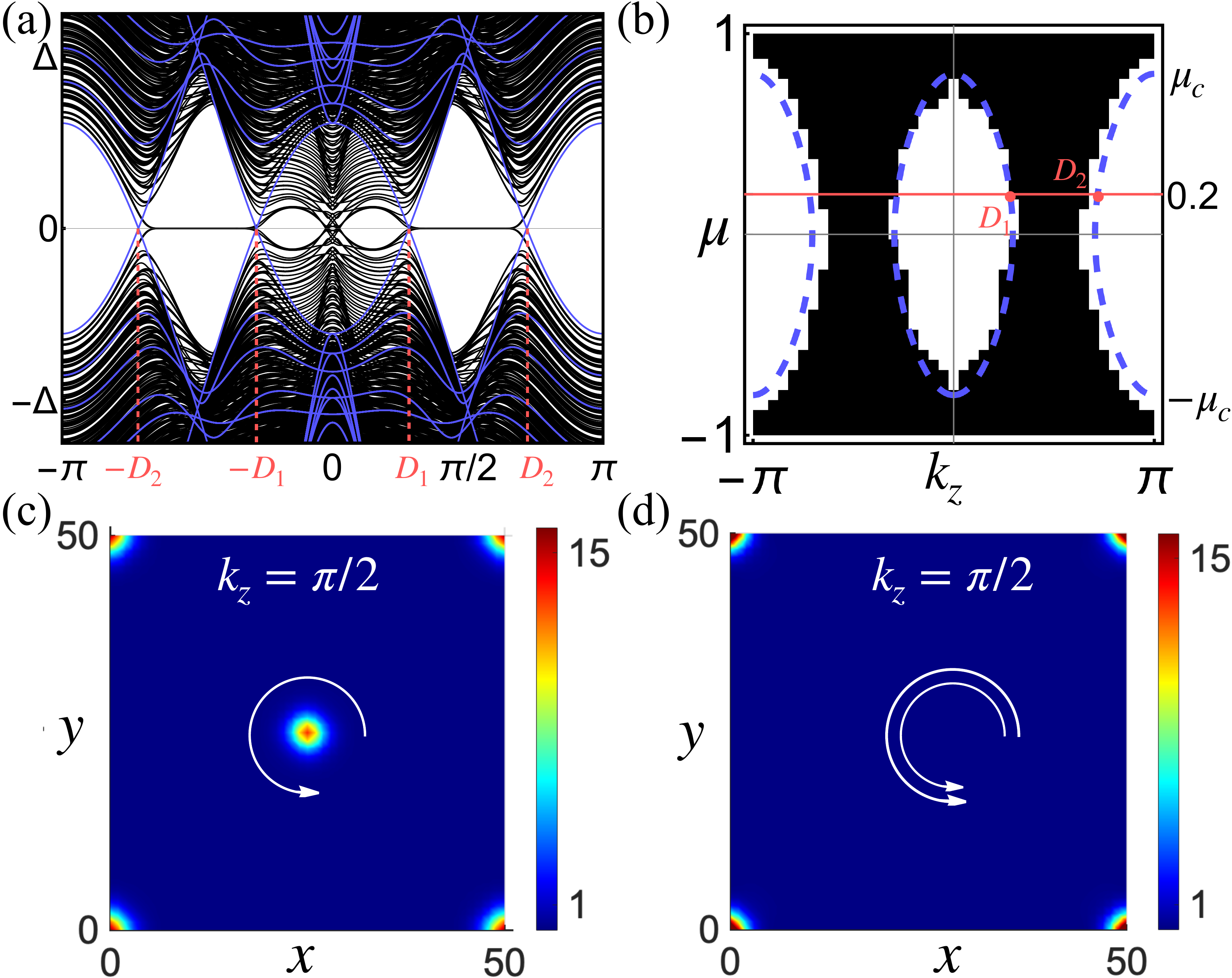}
		\caption{\label{fig_VFBI}
		(a) The black lines show the energy spectrum with a single vortex as a function of $k_z$ when the open boundary conditions apply in $x$ and $y$ directions.
        The blue lines represent the bulk energy spectrum of Eq.~\eqref{eq_bdg} projected to the $\Gamma$-Z line.
        (b) shows the numerical results for the region where vortex flat bands exist in the $(\mu,k_z)$-space. The black region means there exist vortex flat bands. The blue dashed lines are the movement of the BdG Dirac points with the chemical potential $\mu$ and $k_z$. The red line represents the chemical potential used in (a).
        (c) and (d) show the real-space wavefunction profiles of lowest energy state at $k_z=\pi/2$ with vortex winding number 1 and 2, respectively.
        The color bars in (c) and (d) are in the unit of $10^{-2}$.
        The superconducting order parameter is set to be $\Delta_{\text{sc}}=0.6$, and all the other parameters are same as that for Fig.~\ref{fig_band}.
        }
	\end{center}
\end{figure}

Although the paring gap is nodal, the nodes occur on the $k_z$ axis, and we can still consider a vortex line along $z$ axis. This effect can be simulated by introducing a phase winding in the superconducting order parameter, i.e. $\Deltasc(r)=\Deltasc e^{i\theta}$, where $r=\sqrt{\bar{x}^2+\bar{y}^2}$, $\bar{x}=x-x_c$, $\bar{y}=y-y_c$, $(x_c,y_c)$ denotes the vortex core, and $\theta$ is the corresponding polar angle. While the vortex line break the in-plane translational symmetry and time-reversal symmetry, the system still retains translational symmetry along $z$ axis and particle-hole symmetry. Thus, the system is quasi-1D and belongs to the symmetry class D.
For the type-I Dirac semimetal phase, we obtain the energy spectral by diagonalizing the BdG Hamiltonian with open boundary condition along the $x$ and $y$ directions, as shown in Fig~\ref{fig_VFBI}(b). Remarkably, we find six flat bands situated in the region between the projections of BdG Dirac points $\pm D_{1,2}$. The endpoints of flat bands do not exactly coincide with bulk Dirac point due to finite-size effects in the calculations. We further plot the wavefunction of these zero-energy state at $k_z=\pi/2$ on the $x$-$y$ plane, as displayed in Fig.~\ref{fig_VFBI}(c). We observe that these states are localized on hinges and vortex core. According to prior studies~\cite{PhysRevB.106.214510,xie2024hinge}, this $B_{1u}$ pairing hosts four flat-band hinge states in the $k_z$ region between the projections of BdG Dirac points. Therefore, the remaining two flat bands are vortex bound states. This contrast to quasi-1D topological nodal vortex state in Dirac semimetals with an $s$-wave pairing. In the phase diagram shown in Fig.~\ref{fig_VFBI}(b), as the chemical potential $\mu$ moves away from the Dirac point, the region of the vortex flat band gradually expands, eventually spanning the entire 1D Brillouin zone for $|\mu_c| \gtrsim 0.8$. When the winding number of the vortex is even, the double vortex flat bands will be absent, and only four hinge states exist according to the 2D wavefunction of zero-energy states shown in Fig.~\ref{fig_VFBI}(d).

\subsection{Type-II Dirac semiemtal}
\begin{figure}[t]
	\begin{center}
		\includegraphics[width=1\columnwidth]{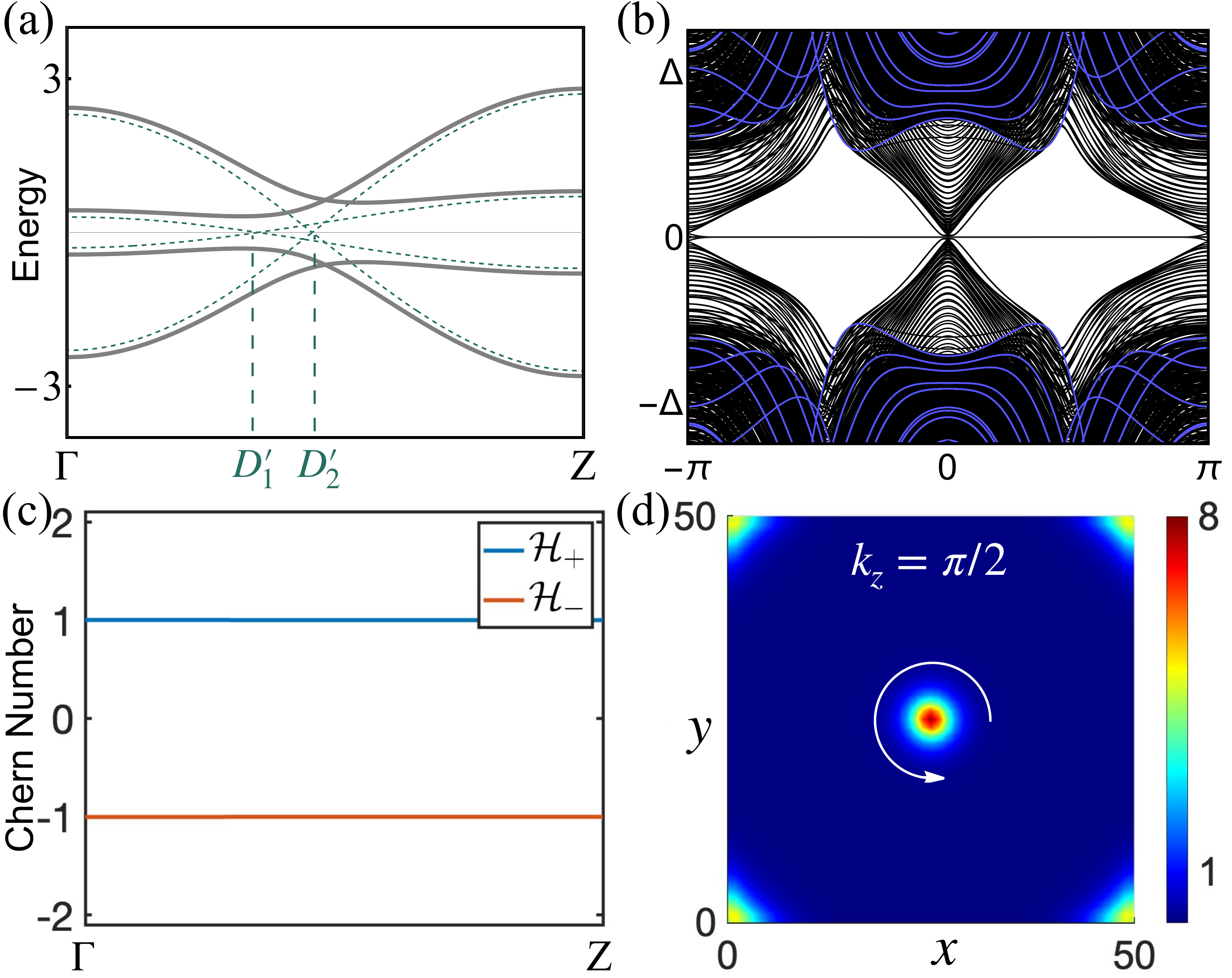}
		\caption{\label{fig_typeII}
        (a) The solid lines show the band structure of normal states described by Eq.~\eqref{eq_bdgII}.
        The dashed lines show the band structure of normal states described by Eq.~\eqref{eq_bdgII} with $\Delta_{\text{sc}}\rightarrow0$.
        (b) The black lines show the energy spectrum with a single vortex as a function of $k_z$ when the open boundary conditions apply in $x$ and $y$ directions.
        The blue lines represent the bulk energy spectrum of Eq.~\eqref{eq_bdgII} projected to the $\Gamma$-Z line.
        (c) The Chern number of $\mathcal{H}_\pm(\kk;k_z)$ as a function of $k_z$.
        (d) show the real-space wavefunction profile of lowest energy state at $k_z=\pi/2$ with a single vortex.
        The color bar is in the unit of $10^{-2}$.
        The parameters are set to be $\{m,t,t^\prime,t_z,\eta,\lambda_1,\lambda_2,\mu,\Delta_{\text{sc}}\}=\{-4,-2,1.5,-1,1.8,1.2,2,0.2,0.6\}$.
        }
	\end{center}
\end{figure}

Additionally, we examine the case of type-II topological Dirac semimetal, where two bands of the Dirac cone have slopes of the same sign along $k_z$. This phase can be achieved by including a titling term and the Hamiltonian is given by~\cite{PhysRevLett.119.026404},
\begin{equation}\label{eq_HII}
    \mathcal{H}_0^{\text{II}}(\kk)=\mathcal{H}_0(\kk)+t^\prime \cos k_z\sigma_0s_0.
\end{equation}
If the second term is sufficiently strong, their slopes of the Dirac cone become both positive or negative, realizing a type-II Dirac semimetal phase.
Accordingly, the BdG Hamiltonian with the same pairing term reads,
\begin{equation}\label{eq_bdgII}
    \mathcal{H}_{\text{BdG}}(\kk)=[\mathcal{H}^{\text{II}}_0(\kk)-\mu]\tau_3+\Delta_{\text{sc}}\sigma_ys_y\tau_1.
\end{equation}
In the limit $\Delta_{\text{sc}} \rightarrow 0$, two pairs of BdG Dirac points are located at $\pm D_{1,2}^\prime$, similar to the type-I case. However, as $\Delta_{\text{sc}}$ increases, the BdG Dirac points at $D_{1,2}^\prime$ can annihilate each other, as illustrated in Fig.~\ref{fig_typeII}(a). At this point, the bulk spectrum of the BdG Hamiltonian becomes fully gapped. Using the same method as before, we can get the energy spectrum with a $\pi$-flux vortex line, as shown in Fig.~\ref{fig_typeII}(b). Intriguingly, six flat bands span the entire 1D Brillouin zone. According to the wavefunctions of zero-energy states displayed in Fig.~\ref{fig_typeII}(d), four of them are high-order hinge states~\cite{xie2024hinge} and two of them are vortex bound states. Therefore, we obtain robust double vortex Majorana flat bands in both type-I and type-II Dirac semimetal phases.

%$k_z$ range, as the Chern number in each subspace remains odd throughout. 

\section{Bulk-vortex Correspondence}\label{sec_topo}
The robust double vortex Majorana flat bands implies a topological origin. In this section, we provide a topological argument for their appearance.
It is convenient to explore the topological invariant in the BdG basis $\Psi_\kk^\prime=(\hat{c}_\kk,\sigma_0s_0\hat{c}_{-\kk}^\dagger)$ and the corresponding Hamiltonian reads,
\begin{equation}\label{eq_HbdgII}
    \begin{split}
\HBdG(\kk)=&\  M(\kk) \tau_z\sigma_z s_0-\mu\tau_z\sigma_0s_0+\Deltasc\tau_y\sigma_ys_0\\
&+\eta (\sin k_x\tau_0\sigma_x s_z-\sin k_y \tau_z\sigma_y s_0)\\
 &+\lambda_1\sin k_z(\cos k_y-\cos k_x)\tau_0\sigma_x s_x\\
 &+\lambda_2\sin k_x\sin k_y \sin k_z\tau_0\sigma_x s_y.
 \end{split}
\end{equation}
Here, $M(\kk)=m-t\cos k_x-t\cos k_y-t_z\cos k_z$. For simplicity, we first ignore the cubic spin-orbit coupling terms and it is apparent that $\HBdG(\kk)$ in Eq.~\eqref{eq_HbdgII} can be decoupled in the spin space, i.e., $\HBdG(\kk)=\mathcal{H}_+(\kk)\oplus\mathcal{H}_-(\kk)$, where
\begin{equation}\label{eq_decoupledH}
\begin{split}
    \mathcal{H}_\pm(\kk)=&\ M(\kk)\tau_z\sigma_z-\mu\tau_z+\Deltasc \tau_y\sigma_y\\
    &\pm\eta \sin k_x \sigma_x-\eta\sin k_y \tau_z\sigma_y.
    \end{split}
\end{equation}
In each subspace $s=\pm$, the residual symmetries $g^\pm$ are represented as
\begin{equation}
    \tilde{\mathcal{I}}^\pm=\tau_z\sigma_z,\  \tilde{\mathcal{M}}_z^\pm=\pm i\tau_0\sigma_0,\  \tilde{C}_{4z}^\pm=\pm i\exp[\pm i\frac{\pi}{4}\tau_z\sigma_z],
\end{equation}
and the particle-hole symmetry takes form $ \tilde{\mathcal{P}}^\pm=\tau_x\mathcal{K}$.

\begin{figure}[t]
	\begin{center}
		\includegraphics[width=0.99\columnwidth]{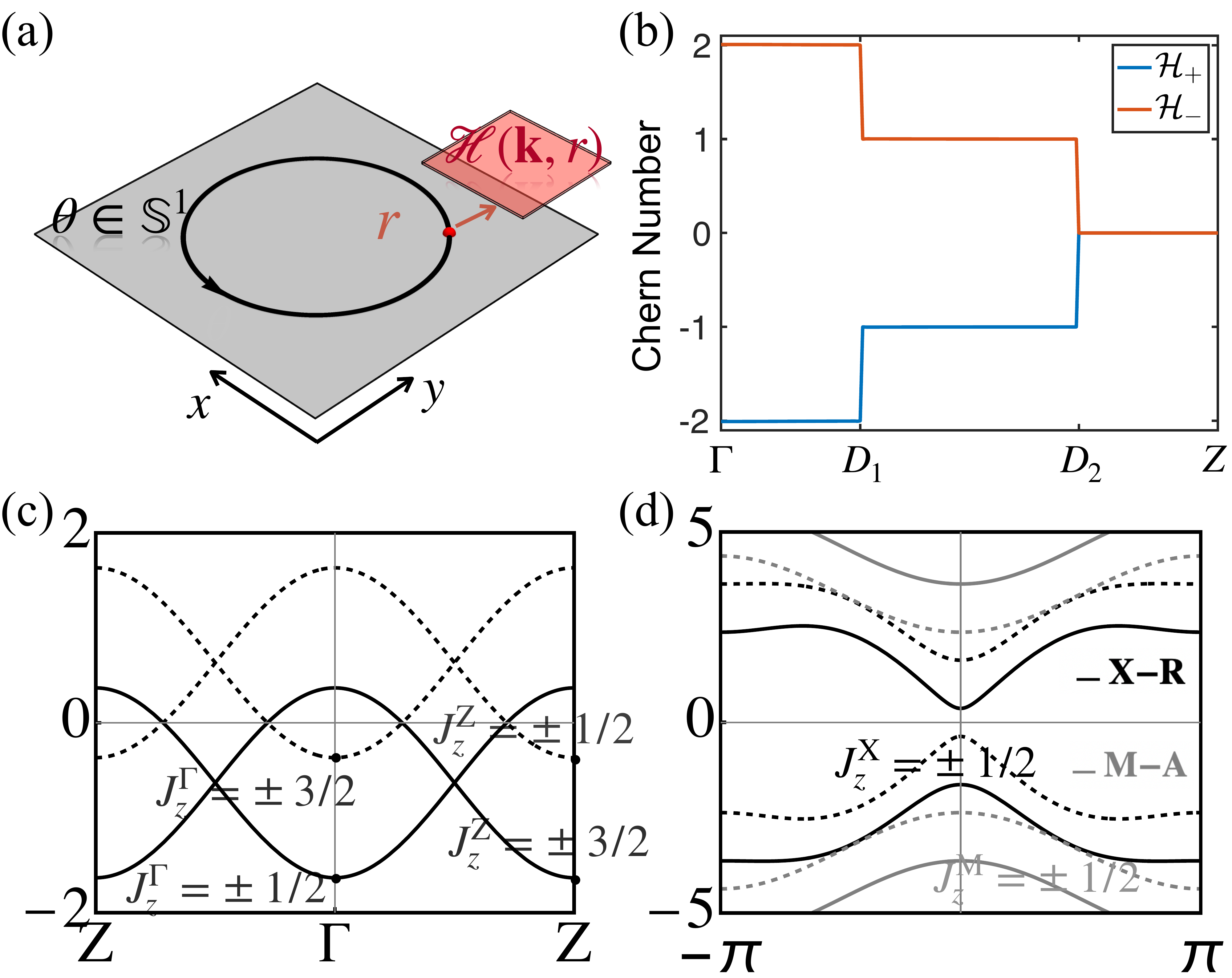}
		\caption{\label{fig_topo}
        (a) The illustration of the base space of target system $\mathcal{H}_\pm(\kk;\theta)$.
        (b) The Chern number of $\mathcal{H}_\pm(\kk;k_z)$ as a function of $k_z$.
        (c) The energy spectrum of $\HBdG(0,0,k_z)$ with the $C_{4z}$ eigenvalues.
        (d) shows the gapped energy spectrum of $\HBdG(\pi,0,k_z)$ (black lines) with the $C_{4z}$ eigenvalues, and $\HBdG(\pi,\pi,k_z)$ (gray lines) with the $C_{2z}$ eigenvalues as a function of $k_z$.
        The solid lines and dashed lines represent electron-like states and hole-like states, respectively.
        All parameters are same as that for Fig.~\ref{fig_VFBI}.
        All labeled angular momenta correspond to negative energy bands.
        }
	\end{center}
\end{figure}

The topological analysis here differs from that in a Dirac semimetal with a quasi-one-dimensional gapless nodal line phase, where the system with a vortex is treated as a 1D system in order to detect 0D invariants in the $k_z=0$ and $k_z=\pi$ planes~\cite{PhysRevLett.123.027003}. To provide a topological perspective on the Majorana vortex flat band, we first consider the Hamiltonian in the $k_z$ plane without a vortex. Except at $k_z = \pm D_{1,2}$, the Hamiltonian $\mathcal{H}_\pm(k_x,k_y;k_z)$ can be viewed as a fully gapped superconductor in class D, with an effective particle-hole symmetry,
\begin{equation}
    \bar{\mathcal{P}}^\pm=\tilde{M}_z^\pm\tilde{\mathcal{P}}^\pm,\ (\bar{\mathcal{P}}^\pm)^2=+\mathbb{1}.
\end{equation}
Next, we consider the BdG Hamiltonian $\widetilde{\mathcal{H}}_{\pm}(\kk;\theta,k_z)$ with inclusion vortex that varies very slowly with the real-space parameter $\theta\in \mathbb{S}^1$ around the vortex, as illustrated in Fig.~\ref{fig_topo}(a).
To achieve this, we can image that the radius of the chosen $\mathbb{S}^1$ is infinitely large.
Now, every point $r$ on $\mathbb{S}^1$ can be considered as a subsystem that is macroscopically small but microscopically large.
This subsystem possesses its own translational invariance, and thus has momentum $\kk\in T^2$.
That is to say, the base space of target system is $T^2\times \mathbb{S}^1$.
Thus, the classification of $ \widetilde{\mathcal{H}}_{\mathrm{BdG}}(\kk; \theta, k_z) $ in each subspace corresponds to the class-D defect classification with defect codimension 2, which is given by $\mathbbm{Z}_2$~\cite{PhysRevB.82.115120,teo2017topological,RevModPhys.88.035005}.
The expression of the $\mathbbm{Z}_2$ invariant is
\begin{equation}
    \nu_\pm=-\frac{1}{4\pi^2}\int_{T^2\times \mathbb{S}^1}\text{Tr}\left[\mathcal{A}_\pm\dd \mathcal{A}_\pm+\frac{2}{3}\mathcal{A}_\pm^3\right]\mod 2,
\end{equation}
where $\mathcal{A}_\pm$ is the Berry connection defined in the subspace $s =\pm$. This expression can be simplified to the product of the Chern number $\mathcal{C}_\pm$ of the 2D subsystem $\mathcal{H}_\pm(\kk)$ and the vortex phase winding number $n$~\cite{PhysRevB.82.115120}:
\begin{equation}
    \nu_\pm=\mathcal{C}_\pm n\mod 2.
\end{equation}
Here, the phase winding number $n$ is 1 for a $\pi$-flux vortex line. The Chern number $\mathcal{C}_\pm$ as a function of $k_z$ is shown in Fig.~\ref{fig_topo}(b). It can be seen that the Chern number changes by 1 as $k_z$ passes through each BdG Dirac point. Consequently, the $k_z$-region between the two BdG Dirac points possesses a nontrivial $\mathbbm{Z}_2$ invariant in each subspace, resulting in stable double-degenerate flat bands. For the case of type-II Dirac semimetal, the Chern number in each subspace remains odd throughout the 1D Brillouin zone ( Fig.~\ref{fig_typeII}(c) ) and thus the system exhibit intact double vortex flat bands. The $\mathbbm{Z}_2$ nature of the vortex flat bands can be further confirmed by vanishing vortex flat bands with an even winding number of vortex (shown in Fig.~\ref{fig_VFBI}(d)). 
Furthermore, the parity of the Chern number $\mathcal{C}_{\pm}$ can be efficiently determined from the symmetry eigenvalues at the high-symmetry points~\cite{PhysRevB.86.115112}: 
\begin{equation}
    (-1)^{\mathcal{C}_{\pm}}=\prod_{n\in\text{neg}}\left[\xi_n^{\pm}(\Gamma)\xi_n^{\pm}(\text{M})\zeta_n^{\pm}(\text{X}) \right]^2,
\end{equation}
where $\xi_n^\pm(\kk)$ is the eigenvalue of $C^\pm_{4z}$ at the $\kk$ point for the negative energy band of $\mathcal{H}_{\pm}(\kk)$, and $\zeta_n^{\pm}(\text{X})$ is the eigenvalue of $C^\pm_{2z}$ at the X point for the negative energy band of $\mathcal{H}_{\pm}(\kk)$.
For conditions in the Fig.~\ref{fig_topo}(b) and (c), we find that $\mathcal{C}_\pm$ are even at $k_z=0$ and $k_z=\pi$.
The BdG Dirac point corresponds to the crossing between the bands with BdG angular momenta $J_z = \pm \frac{1}{2}$ and $J_z = \pm \frac{3}{2}$, which causes a parity change in $\mathcal{C}_\pm$. In other words, the movement of the BdG Dirac points with respect to the chemical potential $\mu$ and the $k_z$ direction forms the boundary of the region where the vortex flat band exists, as indicated by the dashed lines in Fig.~\ref{fig_VFBI}(b).

\begin{figure}[b]
	\begin{center}
		\includegraphics[width=1\columnwidth]{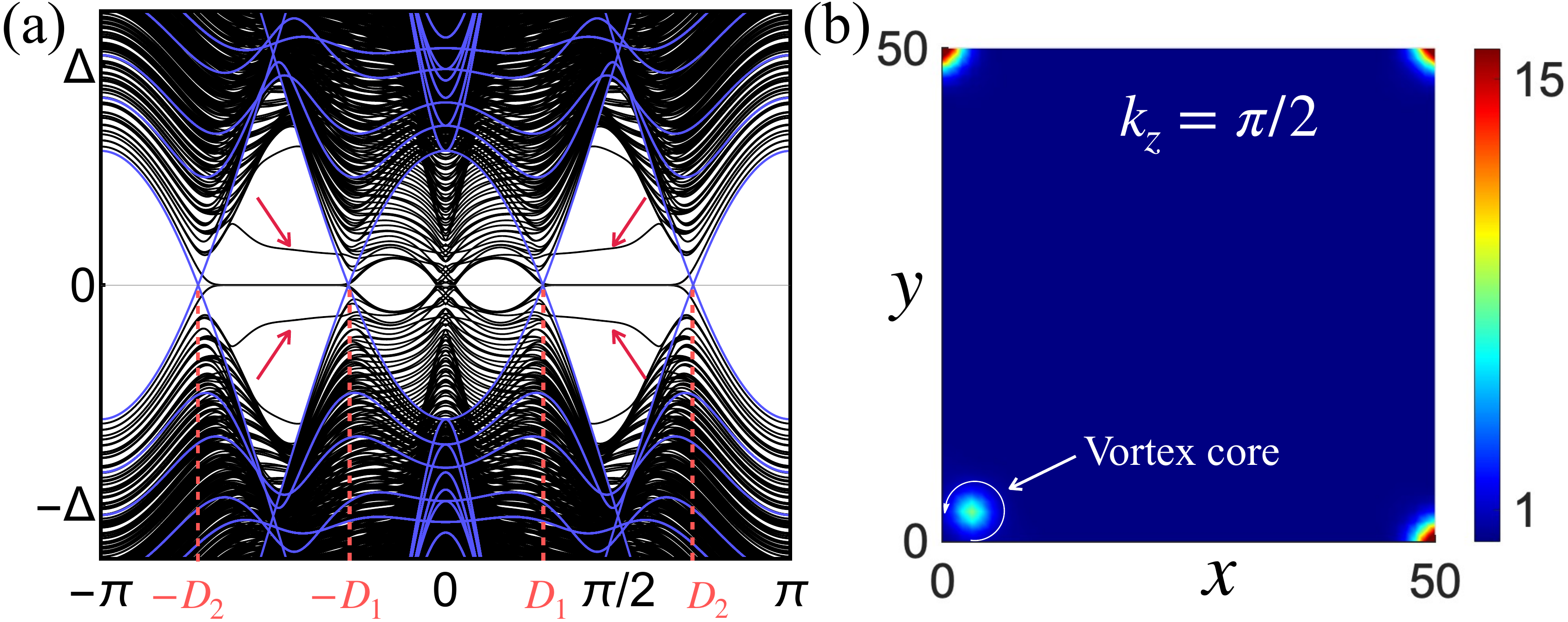}
		\caption{\label{fig_VFBIfaraway}
        (a) The black lines show the energy spectrum with a single vortex located near the boundary,as a function of $k_z$ when the open boundary conditions apply in $x$ and $y$ directions.
        The blue lines represent the bulk energy spectrum of Eq.~\eqref{eq_bdg} projected to the $\Gamma$-Z line.
        The two bands that open a gap due to the coupling between the vortex bound state and the hinge state are marked with red arrows.
        (b) show the real-space wavefunction profiles of lowest energy state at $k_z=\pi/2$ with a single vortex.
        The color bar is in the unit of $10^{-2}$.
        }
	\end{center}
\end{figure}

\begin{figure*}[htbp!]
	\begin{center}
		\includegraphics[width=2\columnwidth]{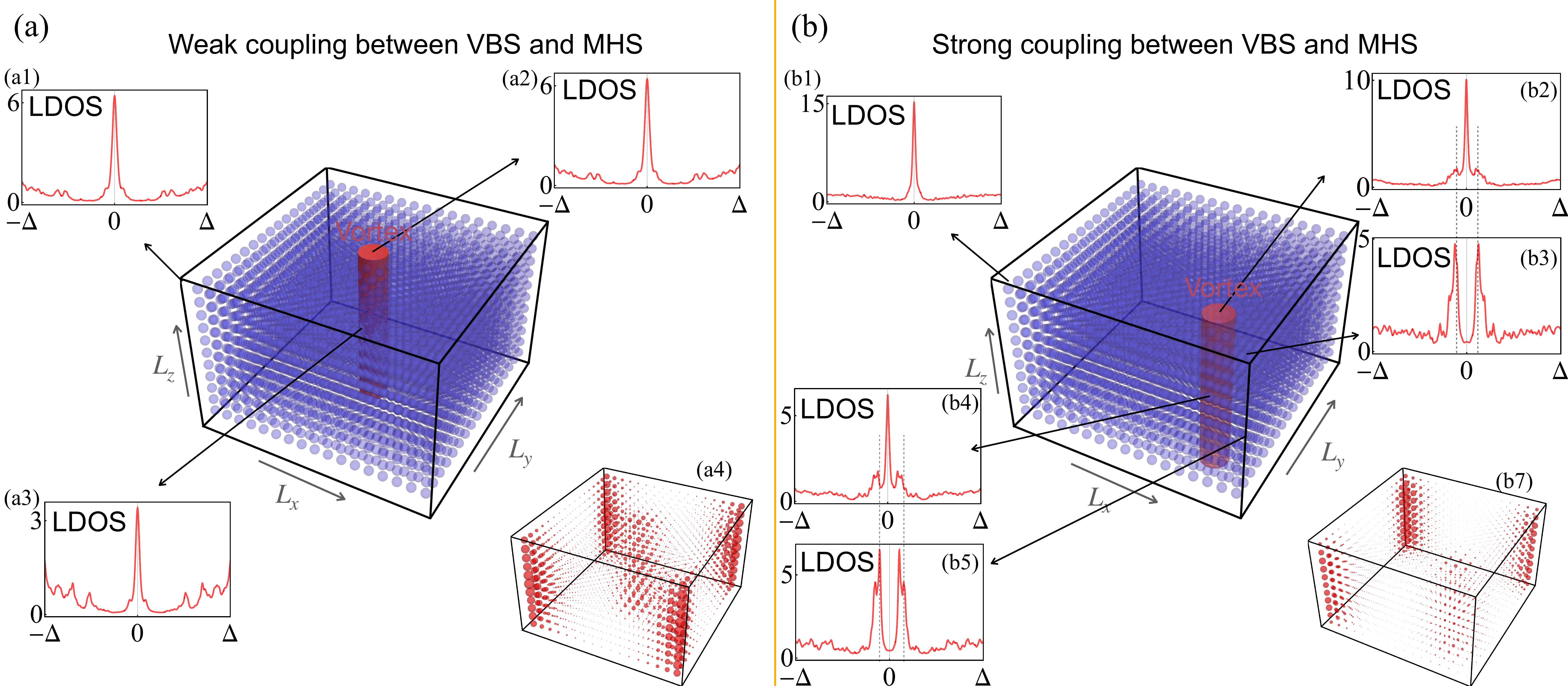}
		\caption{\label{fig_interplay}
        The local density of states and the real-space wavefunction profiles of the lowest energy states for the case where a single vortex is far from each hinge (a), and a single vortex is near one hinge (b).
        All calculations are performed using a cubic lattice, with the lattice size $(L_x,L_y,L_z)=(16,16,10)$, as shown in (a1) and (a2).
        (a1)-(a3) The LDOS calculated at $(l_x,l_y,l_z)=(0,0,10)$, $(8,8,10)$ and $(8,8,5)$, respectively.
        (a4) The real-space wavefunction profiles of lowest energy state. The size of the sphere represents the magnitude of the wavefunction at this lattice site.
        (b1)-(b5) The LDOS calculated at $(l_x,l_y,l_z)=(0,0,10)$, $(13,3,10)$, $(16,0,10)$, $(13,3,5)$ and $(16,0,5)$, respectively.
        The dashed lines mark the same energy in two subfigures.
        (b7) The real-space wavefunction profiles of lowest energy state.
        All the parameters are set to be same as that for Fig.~\ref{fig_VFBI}.
        VBS and MHS mean vortex bound states and Majorana hinge state, respectively.
        }
	\end{center}
\end{figure*}

The double degenerate vortex flat band here is stabilized by the $C_{4z}$ symmetry.
The phenomenon is closely related to the basis choice made during the decoupling in Eq.~\eqref{eq_decoupledH}.
More specifically, we can interpret the orbital degree of freedom as pseudospin degree of freedom, the Hamiltonian Eq.~\eqref{eq_decoupledH} in two spin spaces describes a Rashba electron gas with a Zeeman field $M(\kk)$ with a conventional $s$-wave pairing in the pseudospin space~\cite{PhysRevLett.104.040502,zhang2024topological}.
In contrast to the realistic Rashba electron gas, the angular momenta of bases for $\mathcal{H}_+$ $(\mathcal{H}_-)$ are different and are given by $J_z = +\frac{3}{2},+\frac{1}{2},+\frac{1}{2},+\frac{3}{2}$ $(J_z = -\frac{1}{2},-\frac{3}{2},-\frac{3}{2},-\frac{1}{2})$. Therefore, the conventional-$s$-wave-like term $\Delta_{\mathrm{sc}}\tau_y\sigma_y$ actually inherits the characteristics of $B_{1u}$ pairing symmetry, and carries an angular momentum of 2.
The $C_{4z}$ eigenvalue of the Majorana bound states in the vortex core of the $\mathcal{H}_\pm(\kk,k_z)$ can be calculated by constructing the angular momentum operator,
\begin{equation}
    \hat{\mathcal{L}}_z^\pm=-i\partial_\theta\pm(\frac{1}{2}\sigma_z\tau_z+\sigma_0)+\frac{1}{2}\tau_z,
\end{equation}
where the first term comes from the rotation of the space, the second term is the intrinsic angular momentum coming from orbital basis and Cooper pair with angular momentum 2, and the third term is from the vortex.
For a system depicted by Eq.~\eqref{eq_decoupledH}, each vortex can bind one Majorana zero modes carrying zero angular momentum~\cite{PhysRevLett.104.040502}.
However, the bases here have a global shift in angular momenta compared with the Rashba electron gas case, of which angular momenta are $J_z=+\frac{1}{2},-\frac{1}{2},-\frac{1}{2},+\frac{1}{2}$.
This means that the Majorana bound state in $\mathcal{H}_+$ ($\mathcal{H}_-$) will acquire an angular momentum shift of $1$ $(-1)$ due to the contribution of the second term in $\hat{\mathcal{L}}^\pm_z$.
Thus, the two Majorana vortex bound states live in the different subspaces with angular momenta $L_z=\pm1$, i.e. $C_{4z}$ eigenvalues $\pm i$. In other words, any perturbation that preserves $C_{4z}$ symmetry cannot couple them to open a gap. Therefore, the inclusion of cubic spin-orbit coupling terms $\lambda_{1,2}$, preserving all lattice symmetries, will not destroy Majorana flat bands, further validating our previous topological argument.

\section{Interplay between Vortex Flat Band and Hinge modes}

The orbital-singlet spin-triplet pairing supports both Majorana hinge states and Majorana vortex bound states. Although our calculations focus on the $B_{1u}$ pairing, it is straightforward to show that $B_{2u}$ pairing possesses the same topological property. The existence of Majorana hinge mode can be attributed to the filling anomaly that arises when particle-hole symmetry is neglected~\cite{PhysRevB.106.214510}. In contrast, the double vortex flat bands has a $\mathbbm{Z}_2$ origin from perspective of the topological defects. Since the positions of vortices can vary in real space, we can explore the interplay between Majorana vortex and hinge modes. To study this, we move the vortex line close to one hinge and perform numerical calculations to simulate the interaction. 
 Fig.~\ref{fig_VFBIfaraway}(a) displays the energy spectrum for the case where the vortex line is near the left bottom hinge. We observe that a pair of flat bands open gaps and shifts to non-zero energies, as indicated by red arrows. This gap opening is attributed to the hybridization between double vortex bound states and one hinge state, ultimately leaving one bound state in the vortex core. This is apparent from the 2D wavefunctions of zero-energy states at $k_z=\pi/2$ displayed in Fig.~\ref{fig_VFBIfaraway}(b).

This interplay can also be demonstrated by examining the evolution of local density of states (LDOS) as the vortex moves close to one hinge. We study the LDOS at the vortex core and hinge for two configurations: when vortex is far from hinges and when it is close to one hinge. In Fig.~\ref{fig_interplay}(a), we plot the LDOS of the case where the vortex is far from the boundaries. When the system is open along the $z$-direction, a sharp zero-bias peak appears in the LDOS at each hinge due to the higher-order Majorana flat bands, as shown in Fig.~\ref{fig_interplay}(a1). Similarly, a sharp zero-bias peak also emerges at the vortex core and deeper within the vortex, as illustrated in Figs.~\ref{fig_interplay}(a2) and (a3). This behavior is distinct from the case of topological insulator with superconductivity, where the bulk spectrum is gapped and there is no zero-biased peak in the middle of vortex line. The wavefunctions of zero-energy modes, displayed in Fig.~\ref{fig_interplay}(a4), show that hinge modes has a negligible overlap with vortex-bound states.
When vortex moves close to one hinge, the LDOS of other three hinges remains unchanged, as shown in Fig.~\ref{fig_interplay}(b1). However, for the vortex and the hinge near it, the LDOS undergoes dramatic changes. The couple between vortex bound states and hinge states mentioned above trivialize the hinge states, leading to a vanishing zero-biased peak at this hinge, as shown in Figs.~\ref{fig_interplay}(b3) and (b5). Meanwhile, as seen in Fig.~\ref{fig_interplay}(b2) and (b4), the zero-biased peak is still present at vortex core due to the remaining single Majorana flat band. Additionally, LDOS displays two weak peaks at the energy of the sharp peaks in the LDOS at the hinge owing to Majorana hybridization. The 3D wavefunctions of zero-energy modes are shown in Fig.~\ref{fig_interplay}(b7), where the single Majorana vortex state can be identified.

\section{Discussion and Conclusion}

We further explore the potential experimental implications for superconducting Dirac semimetals~\cite{aggarwal2016unconventional,wang2016observation,he2016pressure}. In conventional Dirac semimetals with weak electronic correlations, superconductivity is likely driven by electron-phonon coupling, resulting in $s$-wave pairing~\cite{PhysRevB.94.014510,PhysRevLett.113.046401,PhysRevLett.115.187001}. However, when electronic interactions are taken into account, interorbital spin-triplet pairing could be favored, as seen in doped topological insulators~\cite{PhysRevLett.105.097001,PhysRevB.90.054503,PhysRevB.90.184512}. Interorbital pairing can be promoted by strong electronic correlation, particularly relevant for IBS.
With electron doping, the Dirac semimetal phase can be achieved in IBS, such as LiFeAs and Fe(Te,Se)~\cite{zhang2019multiple}. If the hole pockets around the $\Gamma$-Z line are absent, the Dirac Fermi surfaces can be isolated around this line, and electronic interactions could drive an orbital-singlet pairing involving \(d_{xz,yz}\) and \(d_{xy}\) orbitals~\cite{PhysRevLett.117.137001,PhysRevB.81.104504}. According to our theory, the orbital-singlet spin-triplet pairing  gives rise to Majorana vortex flat bands and exhibits a sharp zero-bias peak at the vortex core, which can be detected using high-resolution scanning tunneling microscopy~\cite{cite-Machida}. The discussed interplay between vortex bound states and higher-order hinge states can also be experimentally examined. Additionally, the detection of vortex Majorana flat bands provides an effective method for identifying orbital-singlet pairing.

In summary, we investigate the vortex bound states in superconducting Dirac semimetals, focusing on the orbital-singlet spin-triplet pairing state. Intriguingly, we identify double Majorana vortex flat bands that occur in the region between the projections of two BdG Dirac points for type-I Dirac semimetals and extend across the entire 1D Brillouin zone for type-II Dirac semimetals. These double Majorana flat bands originate from a nontrivial $\mathbbm{Z}_2$ topology defined by an effective particle-hole symmetry and are protected by four-fold rotational symmetry. Additionally, we observe that moving the vortex line close to a hinge can trivialize the higher-order Majorana arc on the hinge, leaving a single Majorana mode at the vortex core due to the hybridization of Majorana modes. This single Majorana mode is robust against lattice symmetry breaking. Finally, we discuss potential experimental implications for correlated Dirac semimetals, such as electron-doped IBS.

\section{Acknowledgments}
Z.Z. and H.C.P. acknowledge support from the Croucher Foundation through CIA23SC01 and the Hong Kong Research Grants Council through ECS 26308021.
X.W. is supported by the National Key R$\&$D Program of China (Grant No. 2023YFA1407300) and the National Natural Science Foundation of China (Grants No. 12447103).
S.Q. is supported by the Beijing Institute of Technology Research Fund Program for Young Scholars.
P.Z. was supported by the National Natural Science Foundation of China (No. 12274209), the Fundamental Research Funds for the Central Universities (No. 14380230).
%\newpage

\ \\
\textit{Note added}: After finishing this work, we became aware of an independent work~\cite{zhang2024majorana}, that discusses the single vortex flat band in time-reversal-breaking Weyl semimetals.
In our case, the system is time-reversal invariant, and the vortex flat bands are double degenerate, stabilized by crystalline symmetry.
\bibliography{ref}

\end{document}